\documentclass[12pt]{article}
\usepackage{epsfig}

\def\beq{\begin{equation}}
\def\eeq#1{\label{#1}\end{equation}}
\def\eeqn{\end{equation}}

\def\beqa{\begin{eqnarray}}
\def\eeqa#1{\label{#1}\end{eqnarray}}
\def\eeqan{\end{eqnarray}}

\let\bar=\overbar

\def\Dslash{\not{\hbox{\kern-4pt $D$}}}
\def\dslash{\not{\hbox{\kern-2pt $\del$}}}

\def\msb{{\bar{\ssstyle M \kern -1pt S}}}

\def\Title#1{\begin{center} {\Large {\bf #1} } \end{center}}

\begin{document}

\Title{Technicolor on the Lattice}

\bigskip\bigskip

%+\addtocontents{toc}{{\it D. Reggiano}}
%+\label{ReggianoStart}

\begin{raggedright}  

{\it Claudio Pica, Luigi Del Debbio}\\
SUPA, School of Physics and Astronomy,\\
University of Edinburgh, \\
Edinburgh EH9 3JZ, Scotland\\
~\\
{\it Biagio Lucini, Agostino Patella, Antonio Rago}\\
School of Physical Sciences,\\
Swansea University, \\
Swansea SA2 8PP, UK

\bigskip\bigskip
\end{raggedright}

\section{Introduction}
\label{sec:intro}
The idea of Dynamical Electro--Weak Symmetry Breaking (DEWSB) due to
the existence of a new strongly interacting sector at the TeV scale
beyond the Standard Model of particle physics has been proposed many
years ago~\cite{Weinberg:1975gm,Susskind:1978ms}.  However the
simplest models obtained by a naive rescaling of QCD are inadequate
since they are in contradiction with the experimental evidence of
precision Electro--Weak tests~\cite{Peskin:1990zt,Peskin:1991sw}.
Walking and conformal technicolor theories have been
proposed~\cite{Holdom:1984sk,Yamawaki:1985zg,Appelquist:1986an,Luty:2004ye}
whose dynamics should be sufficiently different from QCD so that they
would not violate the experimental constraints~(for recent reviews of
techicolor models see~\cite{Hill:2002ap,Sannino:2008ha}).  In
particular the idea of using models with matter fields in
representations other than the fundamental near the onset of the
conformal window has been recently
advocated~\cite{Sannino:2004qp,Dietrich:2006cm}.  In this work we
will focus on one of the possible candidate theories, the so-called
Minimal Walking Technicolor theory, based on the gauge group SU(2)
with two Dirac fermions in the adjoint representation.  The attribute
``minimal'' in the name stems from having only 2 flavors and from
being the theory with the estimated minimal value of the
Peskin-Takeuchi S parameter among all theories with matter fields in
one representation only.

Given the difficulty of dealing with strongly interacting theories,
the analytical approaches which led to pinning down the promising
candidates for realistic technicolor theories are all dependent on
uncontrolled approximations or conjectures based on educated
guesses. We approach the problem using Lattice Gauge Theory
(LGT). This is a first principle framework in which quantitative
predictions useful for phenomenology can be obtained, as it has been
extensively demonstrated for QCD.

The lattice comunity has already shown renewed interest in this subject during the last two years~\cite{Catterall:2007yx,Appelquist:2007hu,DelDebbio:2008wb,Shamir:2008pb,Deuzeman:2008sc,DelDebbio:2008zf,Catterall:2008qk,Svetitsky:2008bw,DeGrand:2008dh,Fodor:2008hm,Fodor:2008hn,Deuzeman:2008pf,Deuzeman:2008da,Hietanen:2008vc,Jin:2008rc,DelDebbio:2008tv,DeGrand:2008kx,Fleming:2008gy,Hietanen:2008mr,Appelquist:2009ty,Hietanen:2009az,Deuzeman:2009mh,Fodor:2009nh,DeGrand:2009mt,DeGrand:2009et,Hasenfratz:2009ea,DelDebbio:2009fd,Fodor:2009wk}.
In this work we present a study of the spectrum of the gauge theory
SU(2) with two Dirac adjoint fermions. Given the analytical
uncertainties, it is not clear if this theory lies within the
conformal window or not. Our first goal to understand the model is to
determine if chiral symmetry is spontaneously broken or if the theory
is IR conformal. We look for evidence of the existence of an IR fixed
point by studying the dependence of the mesonic spectrum on the
current quark mass. As explained in detail below, to obtain reliable
results it is essential to control all the well-known sources of
systematic errors introduced by the lattice approximation.
   
In Sect.~\ref{sec:sim} we give an overview on the
treatment of systematic errors present in a lattice computation. 
In Sect.~\ref{sec:res} we present our
latest results and discuss the evidence produced. Our conclusions are
given in Sect.~\ref{sec:concl}.

\section{Lattice simulations}
\label{sec:sim}
Lattice Field Theory reduces the Euclidean path integral to an
ordinary multidimensional integral by approximating the continuum
space-time with a discrete four dimensional hypercubic lattice.  To
guarantee exact gauge invariance, gauge fields are replaced by
parallel transporters $U_\mu(x)$ between neighbouring sites $x$ and
$x+\mu$, while fermion fields $\psi(x)$ live on the lattice sites.
The action for matter fields in a representation $R$ is given
by\footnote{We omit for the sake of simplicity all the position, color
  and spin indexes.}
\begin{equation}
  S(U, \psi, \overline{\psi})=\beta S_g(U) + 
  \sum_{i=1}^{N_f}\overline{\psi_i^R} D_m(U^R) \psi_i^R\,\, ,
\end{equation}
where the exact form of the lattice gauge action $S_g$ and the massive
lattice Dirac operator $D_m$ depend on the particular discretization
used.  For the simplest discretization, the action depends only on two
bare parameters: the bare inverse coupling $\beta$ and the bare quark
mass $am_0$.  While the link variables appearing in the gauge action are
in the fundamental representation of the gauge group, the links in the
lattice Dirac operator are in the same representation $R$ as the
fermions fields.  The partition function, after integrating out the
matter fields, takes the form:
\begin{equation}
  Z = \int \mathrm{exp}[-\beta S_g(U)]  [{\rm det}\ D_m(U^R)]^{N_f} dU \,\, .
\end{equation}

We use the Wilson action in our simulations: the gauge action is
simply proportional to the real part of the trace of 1$\times$1
plaquettes summed over the whole lattice; for the fermions the
Wilson--Dirac operator is given by
\begin{equation}
  D_m(U^R) = am_0 + \frac12 \sum_\mu \left[\gamma_\mu \left(\nabla_\mu + \nabla^*_\mu \right) 
    - a \nabla^*_\mu \nabla_\mu \right]\, ,
\end{equation}
where $\nabla_\mu$ is the discretized forward covariant derivative
depending on the link $U^R_\mu$ and $\nabla^*_\mu$ its adjoint.
In a numerical lattice simulation the computation of the discretized
path integrals is performed by Monte Carlo integration using
importance sampling: an \textit{ensemble} of gauge configurations is
generated with probability proportional to $\exp[-\beta S_g(U)]
[\det\ D_m(U^R)]^{N_f}$.  The expectation value of any observable
can then be computed as a stochastic average over this ensemble of
configurations.

\subsection{Systematic Errors in Lattice Simulations}
%\label{sec:sim}
It is important to realize that the desired continuum results can only
be recovered taking appropriate limits of numerical lattice
simulations.  Changing perspective, one can say that lattice
simulations are always affected by systematic errors, which must be
controlled to obtain reliable results.  This is of the utmost
importance for our calculations, since the theory we aim to studying,
$SU(2)$ with Dirac adjoint fermions, is not understood theoretically
and could be very different from the familiar theory of QCD. We need
to ensure that we are observing genuine features of the continuum
theory, and not artefacts of our lattice formulation. 

Let us remind - and warn! - the reader about the sources of
systematic errors.  Besides statistical uncertainties, arising from the
finite number of configurations in our statistical ensemble, there
are:
\begin{description}
\item[finite-size, finite-temperature corrections]~\\
  Numerical simulations are always performed on a lattice with finite
  extent in all four direction, usually in a geometry
  $T$$\times$$L^3$. On a 4-dimensional torus these
  expectation values can have large corrections, even if the infinite
  volume limit is reached at an exponential rate
  asymptotically
  %\footnote{This is the expected behavior unless we
  %tuned the system to be at a critical point.}
  . The absolute magnitude of finite size effects increases as the
  correlations in the lattice increase, i.e. they become big when the
  Compton lenghtwave of light modes propagating in the lattice is
  comparable with its size. Moreover, if the size of the lattice is
  not sufficiently large, the system may enter a phase that bears
  little resemblance to the large--volume theory we are intersted
  in. Some examples, based on analytical finite--volume
  results~\cite{'tHooft:1979uj,Luscher:1982ma,vanBaal:1986ag}, have
  been recently discussed in Ref.~\cite{Fodor:2009wk}.
\item[explicit breaking of chiral symmetry]~\\
  The lattice discretization breaks chiral symmetry explicitly. For
  Wilson--Dirac fermions chiral symmetry is recovered by tuning one parameter,
 the bare fermion mass, as one decreases the lattice
  spacing. Other formulations are known which possess better chiral
  properties, however all of them are much more computationally
  expensive than Wilson--Dirac fermions\footnote{Kogut-Susskind
    staggered fermions also have better chiral properties than Wilson
    fermions for a comparable cost. However in this case one is
    limited to having a number of flavors which is a multiple of four,
    in order to avoid complications due to the so--called rooting of
    the fermion determinant.}.  Even with better chiral formulations,
  there is another technical problem: the zero modes of the lattice
  Dirac operator, which appear as the chiral limit is approached, make
  it impossible to run simulations at very light masses with the
  current algorithms. When using Wilson fermions on a given finite
  lattice there is a lower limit to the masses which are numerically
  accessible, which depends on the volume of the system.
\item[lattice artefacts]~\\
  Computations are performed on a discrete four dimensional
  lattice. Since we are ultimately interested in the continuum
  physics, we must consider the limit in which the lattice spacing,
  conventionally called $a$, goes to zero. However the lattice spacing
  $a$ is not a parameter of the simulation that one can directly
  change (all quantities in a simulation are dimensionless). The
  continuum limit is recovered in a more subtle way: approaching the
  UV fixed point, at bare coupling $g=0$, i.e. $\beta=\infty$, the
  system undergoes a continuous phase transition and, by universality,
  the microscopic details of the system becomes immaterial. To put it
  in a different way, the physics at the scale of the lattice spacing
  must decouple from the long range (continuum) physics which we are
  interested to study on the lattice.
\end{description}

To give reliable predictions, lattice
simulations must be performed in a regime in which a precise hierarchy
of scales is realized.  To illustrate this point let us consider first
the case of QCD.  Numerical simulations, as explained above, are always
performed with a finite quark mass, which explicitly induces a mass
gap in the theory. To avoid finite-size effects in the computation of
the low-lying spectrum, the lattice size $L$ must be much bigger than
the inverse mass of the lightest hadron $m_\mathrm{PS}$ we are trying to
measure. Since we are interested in the chiral regime, this light
hadron mass must be much smaller than the characteristic hadronic
scale, at which the theory becomes strongly interacting: we can take
for example the Sommer scale $r_0$, as a convenient quantity to
measure on the lattice.  Finally to avoid discretization errors, the
lattice spacing must be much smaller than the reference scale for
strong interactions $r_0$, so that the physics at the scale of the UV
cutoff is weakly interacting.  In summary we must have the following
hierarchy of scales:
\begin{equation}
  \left( \frac{L}{a} \right)^{-1} \ll a m_\mathrm{PS} \ll 
  \left( \frac{r_0}{a} \right)^{-1} \ll 1\,\, ,
\end{equation} 
for the computations to be reliable, and only in this regime we can
safely extrapolate to the chiral and continuum limit.

Let us now assume that the continuum theory we are interested in
studying on the lattice has an IR fixed point in the massless
limit. Introducing a mass term, as it is always the case in numerical
simulations, explicitely breaks conformal invariance and the above
considerations made for QCD remain valid also in this case.  In
particular it is still possible to define a scale in analogy to $r_0$
in QCD (see Sect.~\ref{sec:disc} below) which remains finite also in
the massless limit. This scale however has a different interpretation
than in QCD.  The difference with the familiar QCD-like case is in the
behavior of the two theories approaching the chiral limit. We will
discuss below in Section~\ref{sec:res} the signatures of the IR
conformal case.

\subsection{Simulation Code}
%\label{sec:code}
Our results presented in the next Section are obtained using our own
simulation code, written from scratch for the specific purpose of
studying gauge theories with fermions in arbitrary
representations~\cite{DelDebbio:2008zf}.  This code was designed to be
flexible, fast, easy to use and extend.
Our simulation code is suitable for the study of gauge theories with:
\begin{itemize}
\item gauge group $SU(N)$ for any $N$;
\item generic fermion representations. At present the code implements
  the fundamental (fund), adjoint (ADJ), 2-index symmetric (S) and
  antisymmetric (AS). It is easy to extend the code to other
  representations, like 3-index symmetric for example, and even to
  have fermions in two or more different representations at the same
  time. In particular no modifications for the computation of the HMC
  force are needed, which is typically the most complicated part of the
  code;
\item any number of flavors. We use Wilson fermions and the RHMC
  algorithm, which is an exact algorithm for any number of flavors.
\end{itemize}
Since this was a new code, and lattice simulations in the past were
mainly devoted only to QCD, we made a large effort to validate the
code and to study its behavior in the parameter region of interest for
physics studies. As part of this effort, we made a number of
cross-checks:
\begin{itemize}  
\item for $SU(3)$, $N_f=2$ with different, established codes (we used
  M.~Luscher's DD-HMC algorithm~\cite{Luscher:2005rx});
\item consistency among different representations (e.g. $SU(3)$ AS vs
  fund, $SU(2)$ SYM vs ADJ);
\item observables with different codes (e.g. meson spectrum for
  $SU(2)$ ADJ in Chroma);
\item large quark mass limit compared to pure gauge (quenched)
  spectrum;
\item correctness of integrator, reversibility, acceptance
  probability;
\item independence on integrator step size.
\end{itemize}
To control the stability of the simulations, which could incur in
well-known problems close to the chiral limit~\cite{DelDebbio:2005qa},
we monitored the lowest eigenvalue of the (pre-conditioned)
Wilson--Dirac matrix. We found no instabilities in our runs. A
detailed report will appear elsewhere~\cite{Pica:2009}.

\section{SU(2) with 2 Adjoint fermions}
\label{sec:res}
Before looking at the actual numerical results, we will discuss the
signatures of an IR fixed point. In this work, we search for
indications of IR conformal behavior in the spectrum. 
%This is not the
%only possible way as one can, for example, study the non-perturbative
%running of a coupling defined in some particular scheme. In fact most
%of the claims of the existence of IR fixed points so far have been
%made by looking at the evolution of the coupling in the Schr\"odinger
%functional
%scheme~\cite{Appelquist:2007hu,Shamir:2008pb,Hietanen:2009az},
%although no solid indications have emerged yet.
%We will briefly discuss later in Section~\ref{sec:disc} on the status of such claims.
If there is no IR fixed point, the theory is expected to be confining
and chiral symmetry to be spontaneously broken. 
This is the familiar scenario of QCD:
% in the chiral limit the
%pseudo-scalar particles (pions) become massless, while the other
%states in the spectrum remain massive; the theory has a non-zero
%pseudoscalar decay constant $F_\mathrm{PS}$ and chiral condensate
%$\langle\overline\psi\psi\rangle$.  
using the PCAC mass from the axial
Ward identity, $m$, to parametrize the explicit breaking of chiral
symmetry, we expect the usual scaling $m^2_\mathrm{PS}\propto m$, as
$m\rightarrow 0$;  $F_\mathrm{PS}$ remains finite; 
the Gell-Mann--Oakes--Renner relation is satisfied
and can be used to extract the non-zero chiral condensate: $(m_\mathrm{PS}
F_\mathrm{PS})^2/m \rightarrow -\langle\overline\psi\psi\rangle$.

On the other hand, if the theory has an IR fixed point, 
and as before,
we parame\-trize the explicit breaking of chiral and conformal symmetry with $m$, 
as $m\rightarrow 0$ we expect that:
\begin{itemize}
\item all hadron masses vanish proportionally to the same power of
  $m$: $m_\mathrm{had} \sim m^{\alpha}$;
\item in particular the ratio of the vector to pseudoscalar meson mass
  remains finite: $m_\mathrm{V}/m_\mathrm{PS}\rightarrow \mathrm{const}<\infty$;
\item $F_\mathrm{PS}$ and $\langle\overline\psi\psi\rangle$
  $\rightarrow 0$.
\end{itemize}
Furthermore, if one measures the string tension $\sigma$ as a function
of $m$, we expect that $\sigma$ vanishes in the chiral limit.

The behavior just described assumes that the system is in the
continuum limit and in an infinite volume.  However this is not the
case in a lattice simulation. In particular the finite size of the
system can be seen as a relevant coupling which will drive the system
away from criticality under the renormalization group (RG) flow.  As
$m$ is decreased towards the chiral limit, great care must be taken to
control finite size effects as explained in Section~\ref{sec:sim}. We
will show below that these effects are quite big in the range of
masses and lattice volumes which are currently used in numerical
simulations.  In particular it is important to keep in mind that an
asymptotically free theory will look almost conformal in a very small
volume and practical limitations can make it very difficult to
distinguish between the two cases.

\subsection{Spectrum}
\label{sec:spec}

We now present the results of our numerical work.  All the simulations
used in this work are made at a fixed lattice spacing, corresponding
to a bare coupling $\beta=2.25$. This value of the coupling was chosen
based on previous studies of the same
theory~\cite{Catterall:2007yx,Catterall:2008qk,Hietanen:2008mr} to
avoid a bulk phase transition present at about $\beta=2.0$.  
%The
%extrapolation towards the continuum limit requires a new series of
%runs at different values of the bare coupling, and is left to future
%investigations.
 
We use four different lattices: $16\times 8^3$, $24\times 12^3$,
$32\times 16^3$ and $64\times 24^3$.  For each of these four lattices
a number of ensembles corresponding to different quark masses were
generated focusing in the range corresponding to pseudoscalar masses
of 0.6--0.2 $a^{-1}$. 
%As explained above, when using Wilson fermions,
%the chiral and infinite volume limits are intertwined. In practice for
%the simulation to be stable, one cannot arbitrarily decrease the quark
%mass without also increasing the volume. This is also necessary to
%keep under control finite-volume effects thus remaining in the
%so-called $p$-regime.  
We explicitly control finite-size effects making simulations with different volumes.  In this
way we can reach the small mass region and study the finite-volume
effects for the first time in this theory. All the previous studies
cited above~\cite{Catterall:2007yx,Catterall:2008qk,Hietanen:2008mr}
have almost no information on this small quark mass region, having
mostly pseudoscalar masses of order of the ultraviolet cutoff
$a^{-1}$ or heavier and no study of finite volume effects.
For each lattice and quark mass we accumulated a statistical ensemble
of about 5000 thermalized configurations, except at the largest volume
for which we present only preliminary data based on approximately 400
thermalized configurations for each quark mass. The gauge configurations were
generated using trajectories in the molecular dynamics integration of
length 1 for the two smallest volumes and 1.5 for the two largest
lattices, with integration parameters leading to an acceptance rate of
about 85\% in all cases, and to an integrated autocorrelation times for
the lowest eigenvalue of the Wilson--Dirac operator of order 10 or
less.

For each ensemble of configurations, we measured the quark mass from
the axial Ward identity (PCAC mass) $m$, the pseudoscalar meson mass
$m_\mathrm{PS}$, the vector mass $m_\mathrm{V}$ and the pseudoscalar decay
constant $F_\mathrm{PS}$.
\begin{figure}[tp]
\begin{center}
\epsfig{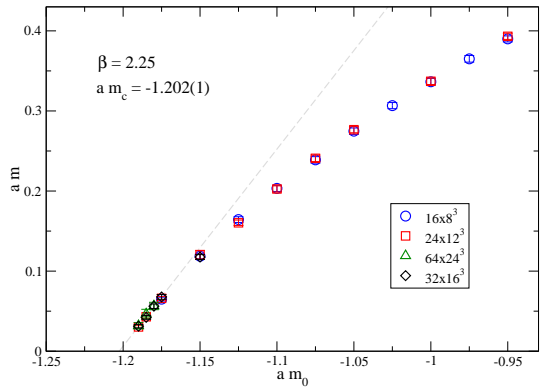}
\caption{Extrapolation of the quark mass from the axial Ward identity
  to locate the chiral limit. As expected no significant finite size
  effects are present.\label{fig:cl}}
\end{center}
\end{figure}
We locate the chiral limit at the critical bare mass where the PCAC
mass vanishes. This does not correspond to bare mass zero because the
explicit breaking of chiral symmetry with Wilson fermions induces an
additive renormalization of the quark mass.  We show in
Fig.~\ref{fig:cl} the extrapolation of $m$ for different lattice
sizes. Using a linear extrapolation, the chiral limit can be located
at the critical bare mass $am_c=-1.202(1)$. As expected from the fact
that $m$ is essentially an UV quantity, no significant finite-size
effects are visible and the measured values for this quantity agree
within errors on all four lattices.
% In the following the PCAC mass will be used instead of the bare
% quark mass as a more convenient quantity.

\begin{figure}[tp]
\begin{center}
\epsfig{file=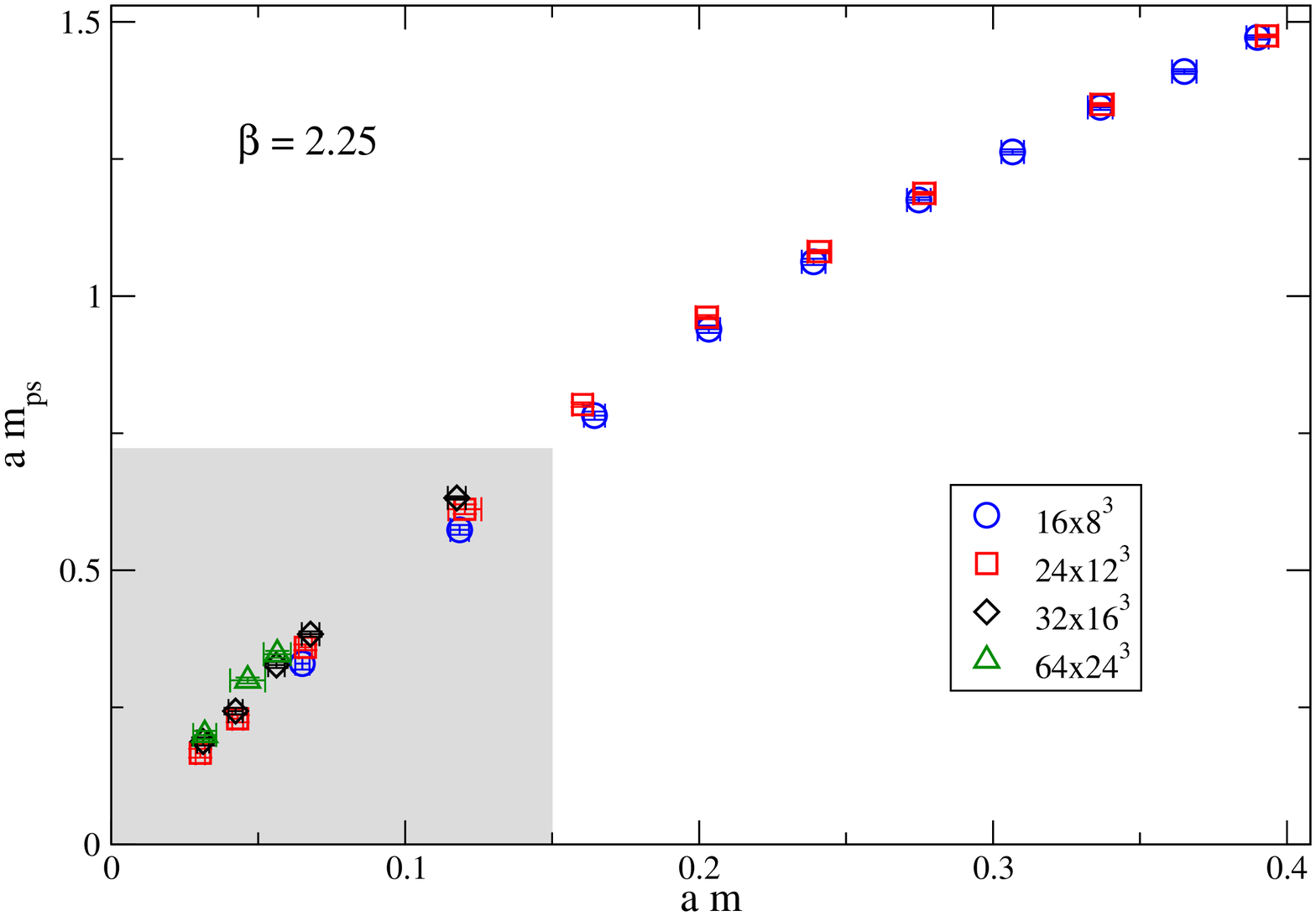,height=2.0in}\hfill
\epsfig{file=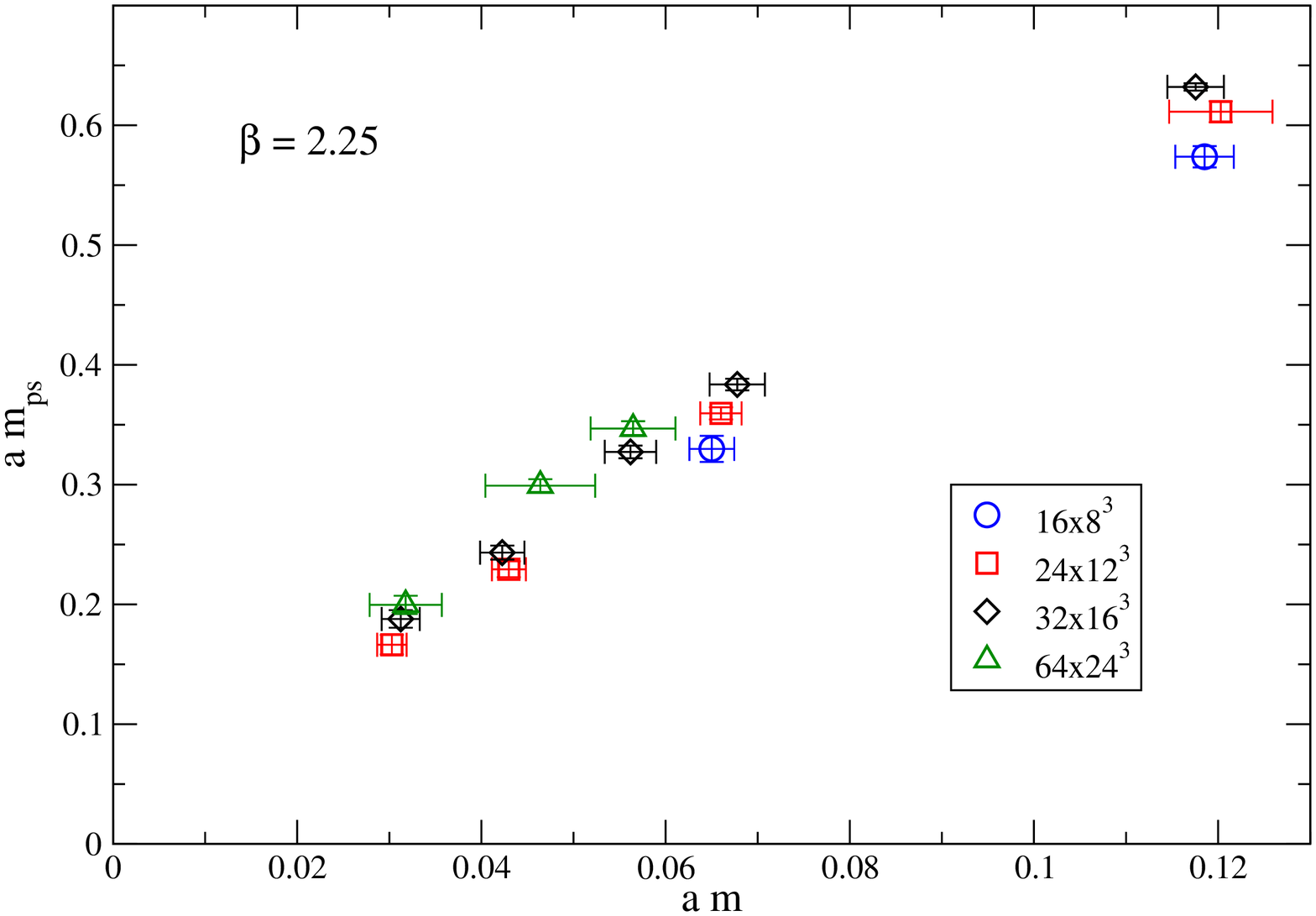,height=2.0in}
\caption{Pseudoscalar meson mass as a function of the PCAC mass. The
  interesting small mass region shaded in the left panel is enlarged
  on the right. Finite volume effects are evident and grow approaching
  the chiral limit.\label{fig:ps}}
\end{center}
\end{figure}
Our results for the mass of the pseudoscalar meson are presented in
Fig.~\ref{fig:ps}. The interesting region of small quark mass is
enlarged in the right panel. The data presented for the $64\times
24^3$ are still preliminary and have large statistical uncertainties.
Finite size effects on $m_\mathrm{PS}$ are evident and, as the PCAC
mass is decreased, they increase as expected from the discussion of
Sect.~\ref{sec:sim}.  To quantify this systematic effect and to keep
it under control, we use larger lattices as the chiral limit is
approached.

This large finite size effects are clearly visible in the range of
volumes and quark masses explored.  Due to this effect it is difficult
to draw definite conclusions about the functional behavior of the
pseudoscalar mass in the chiral limit. At the current level of
accuracy, it is not possible to exclude a QCD-like behavior for the
pseudoscalar mass, as well as for all the other quantities we measure,
as it will be clear in the following. Data at a finite volume tend to
suggest an IR conformal behavior, but it is not clear whether or not
this behavior will persist once the infinite volume limit is taken
before the chiral limit.

\begin{figure}[tp]
\begin{center}
\epsfig{file=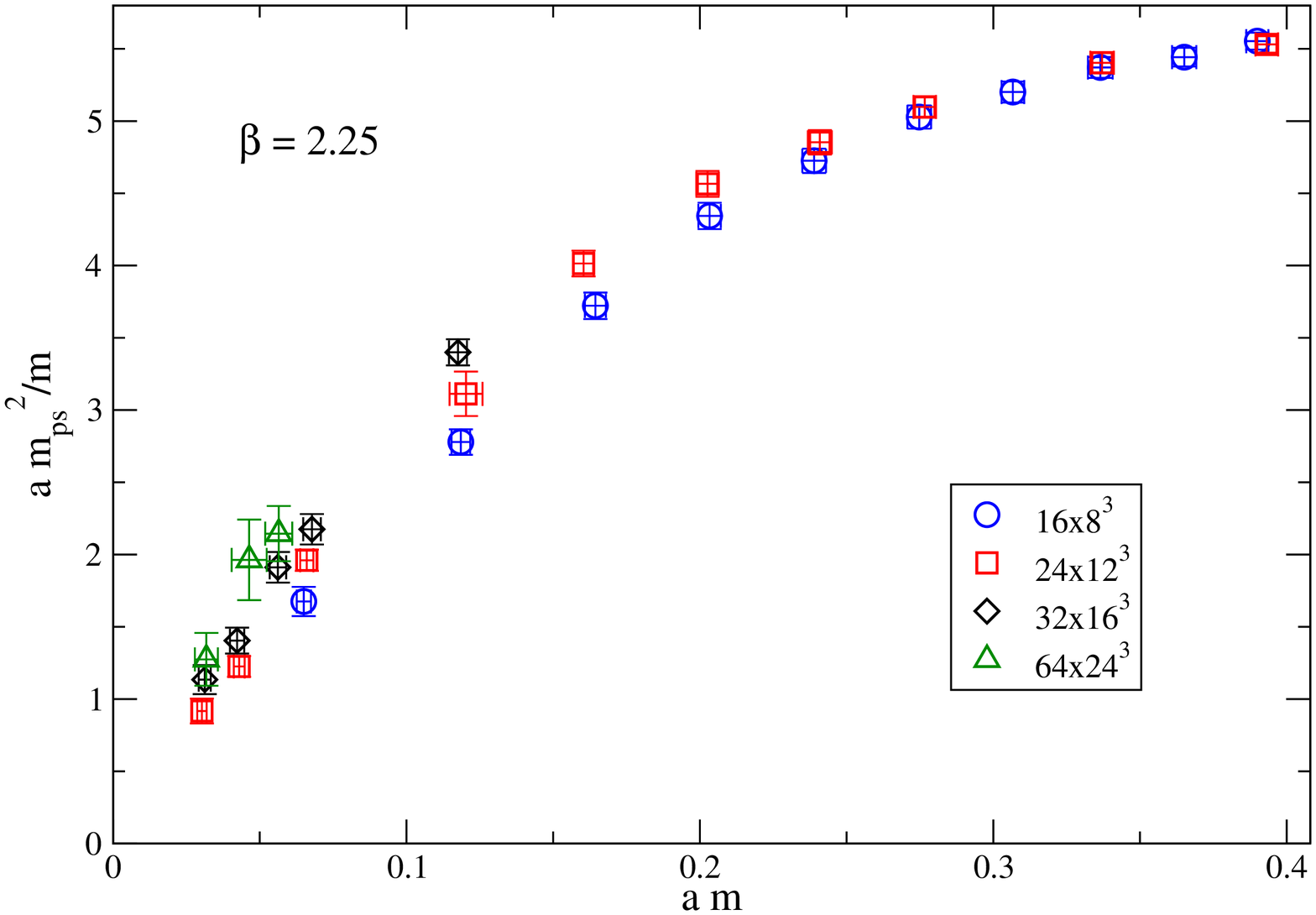,height=2.0in}\hfill
\epsfig{file=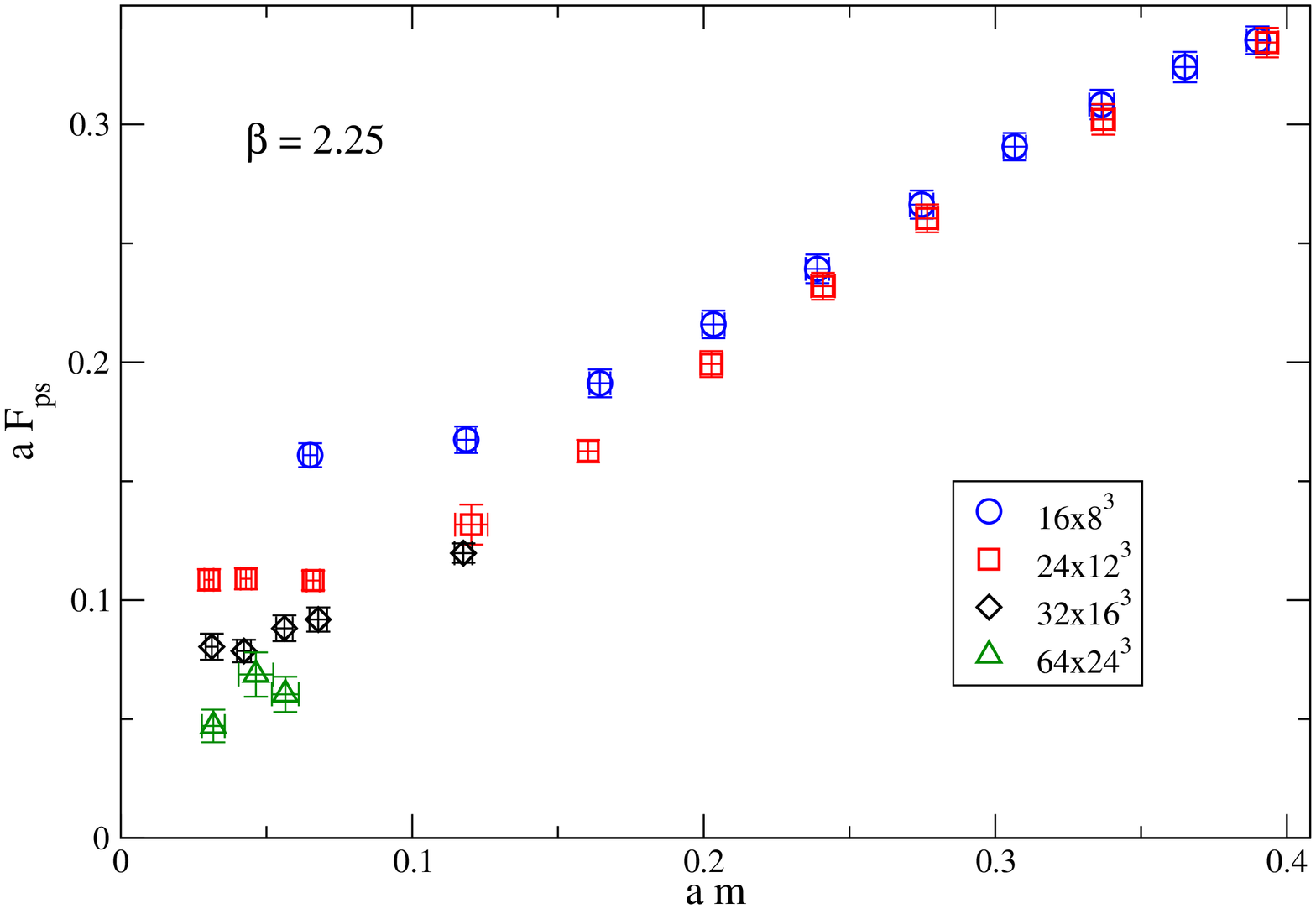,height=2.0in}
\caption{(Left) Ratio of the pseudoscalar mass squared to the PCAC mass. The
  extrapolation to the chiral limit suffers from large finite-volume
  effects. See the text for a discussion. (Right) Pseudoscalar decay constant near the chiral limit. Very large
  finite volume effects are present also in this case which cause the
  chiral extrapolation to be have large uncertainties.\label{fig:ratio}}
\end{center}
\end{figure}
To exemplify this behavior which, as mentioned, is common to all the
quantities we measure, consider the ratio $am_\mathrm{PS}^2/m$ shown
in Fig.~\ref{fig:ratio} (left).  If the theory has an IR fixed point the
ratio should vanish in the chiral limit, while it remains finite in
the other case.  The data appear to be well described by a linear
function of the PCAC mass for $am \leq 0.15$. If we consider our two
smallest volumes, the extrapolated value is quite small and compatible
with zero within errors. As we move to larger volumes however, the
meson masses increase and the extrapolated value starts to deviate
from zero, while remaining very small.
%The present level of accuracy doesn't allow to reach a conclusive answer.
% We will discuss below in Sect.~\ref{sec:disc} the two possible scenarios and 

%\begin{figure}[tp]
%\begin{center}
%\epsfig{file=fpi.eps,height=2.0in}
%\caption{Pseudoscalar decay constant near the chiral limit. Very large
%  finite volume effects are present also in this case which cause the
%  chiral extraplation to be have large uncertainties.\label{fig:fps}}
%\end{center}
%\end{figure}
We consider next another physically interesting quantity, the
pseudoscalar decay constant $F_\mathrm{PS}$, shown in
Fig.~\ref{fig:ratio} (right).  The chiral extrapolation is problematic and not
conclusive also in this case due to large finite volume effects. This
quantity however behaves in the opposite way than the pseudoscalar
mass $m_\mathrm{PS}$. At finite volume the chiral extrapolation yields
a finite result: $F_\mathrm{PS}$ reaches a plateau for masses smaller
than some volume--dependent threshold value of the PCAC mass. However
as the volume is increased, the plateau value of $F_\mathrm{PS}$
decreases and a clear envelope for the curves at different volumes is
visible in Fig.~\ref{fig:ratio} (right).  An extrapolation of this envelope to
the chiral limit yields small numerical values, with large systematics
due to choice of the range used in the fit.  It is also interesting to
note that the plateau values decrease approximatively like $1/L$, as
one should expect in the conformal case.

\begin{figure*}[t]
\begin{center}
\epsfig{file=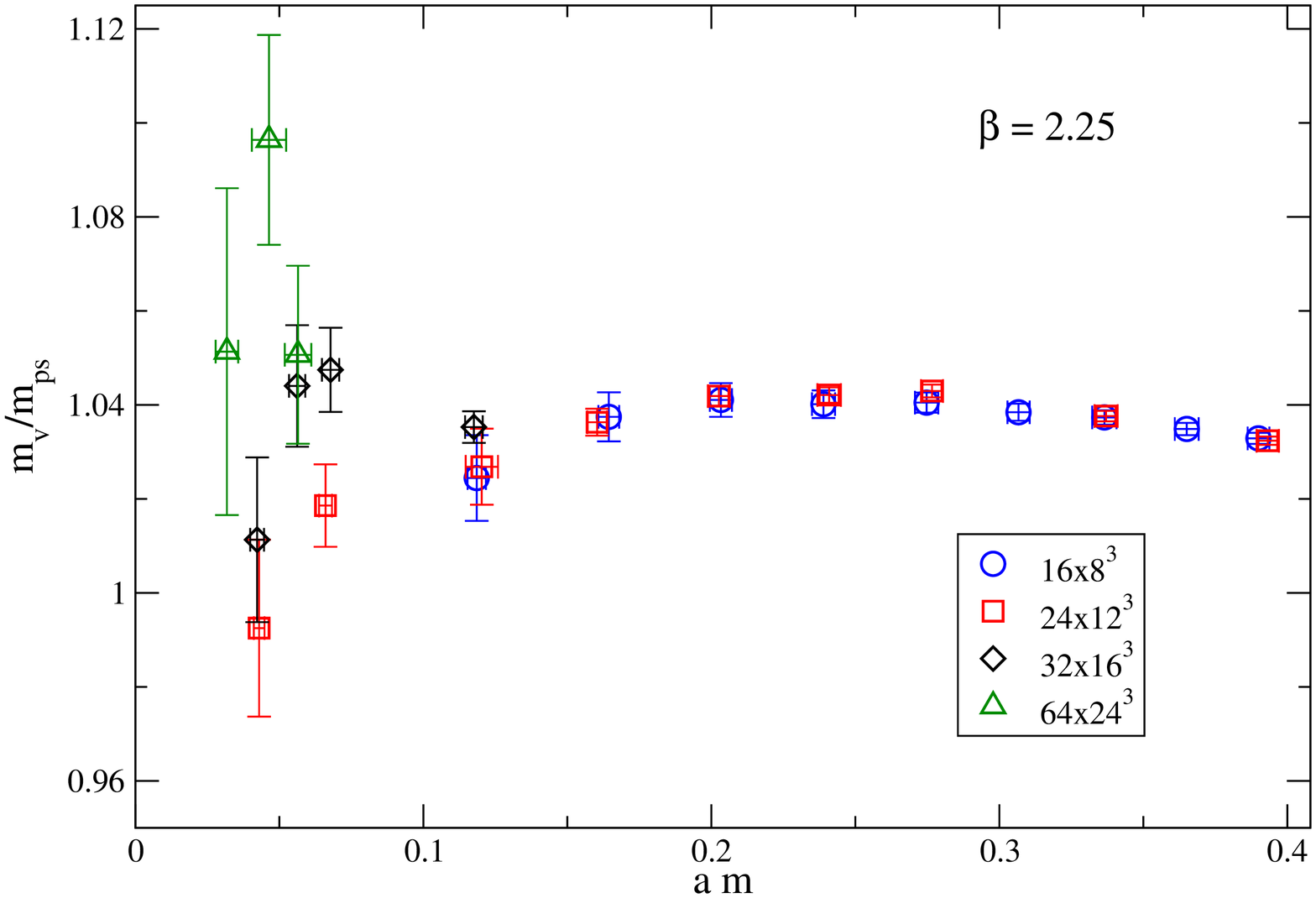,height=2.0in}\hfill
\epsfig{file=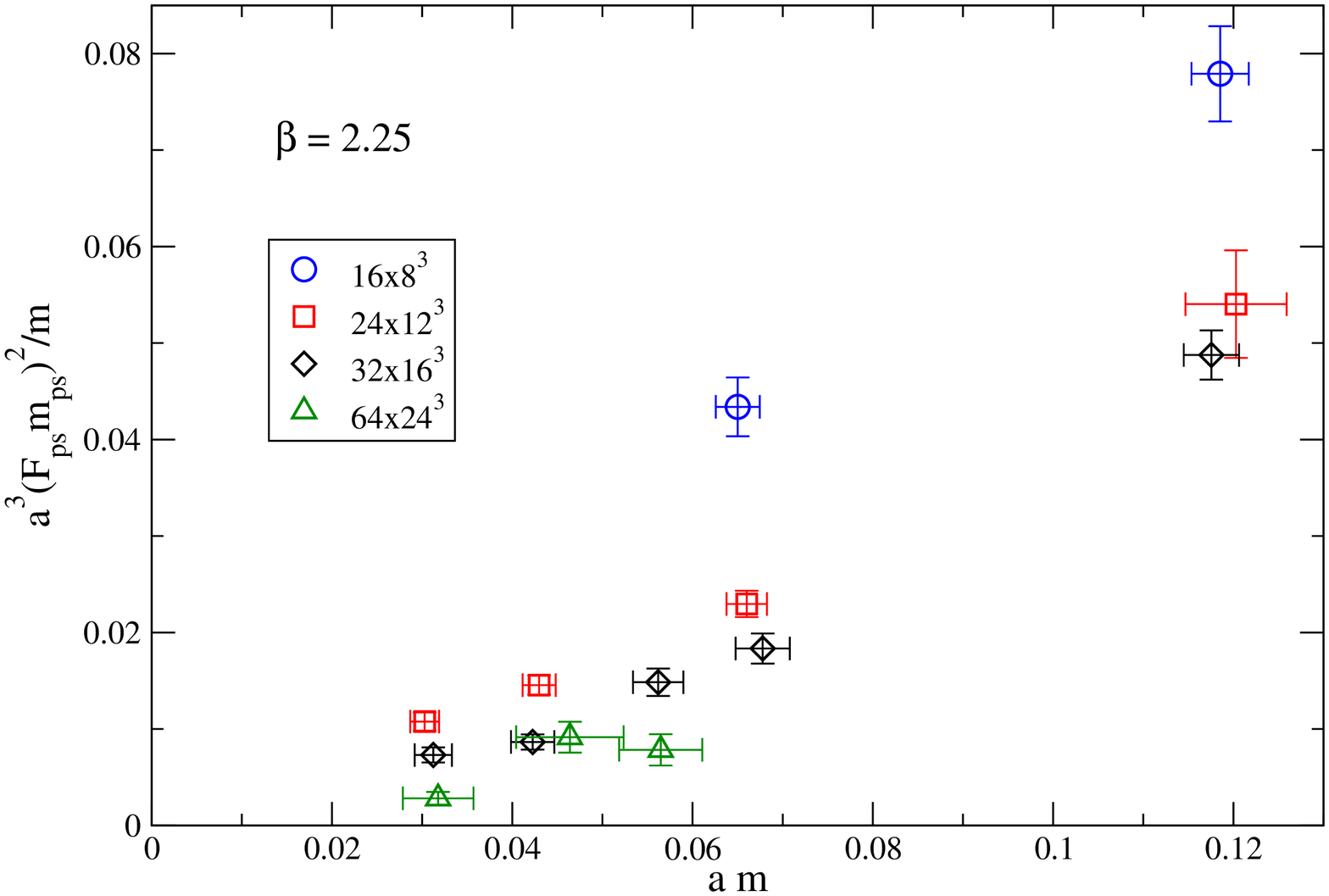,height=2.0in}
\caption{(Left) Comparison between the vector and pseudoscalar meson
  masses. At large PCAC mass, due to quenching the ratio is very near
  to one. Near the chiral limit large finite size effects show
  up. (Right) The GMOR relation can be used to extract information on the
  chiral condesate. The measure results however quite difficult in
  practice and we cannot distinguish any signal of chiral symmetry
  breaking.\label{fig:mvomps}}
\end{center}
\end{figure*}
The ratio of the vector to pseudoscalar mass is shown in
Fig.~\ref{fig:mvomps} (left). This quantity is bounded to be greater than
one~\cite{Weingarten:1983uj} and in the heavy quark limit will tend to
one.
% \footnote{Due to lattices wth finite extent in the temporal
%   direction and systematic uncertainties in the extraction of the
%   mesonic masses, it can however appear to be less than one in an
%   actual simulation.}.
At large $m$ finite volume effects are small, the ratio is bigger than
1, although deviating only by 4\%, and decreasing as $m$ increases, as
expected in the heavy quark approximation.
% A more detailed comparison with the quenched theory is possible
% which we~\cite{DelDebbio:2009fd}.
It should be clear that the region $am>0.2$, which corresponds to
mesons with masses of order of the inverse lattice spacing or more, is
not suitable for studying the physics of the continuum, and that in
the same region the theory is effectively quenched. Nevertheless what
is remarkable in Fig.~\ref{fig:mvomps} (left) is the fact that in the whole
mass range we were able to explore, no large separation between the
pseudoscalar and vector mass is observed.
Looking at our data in more detail, we observe that at fixed volume
the ratio $m_\mathrm{V}/m_\mathrm{PS}$ presents a maximum for some
value of $m$ below which it becomes an increasing function of
$m$. Absolute deviations from the maximum value $\approx 1.04$ are
small, and one may be tempted to conclude that this ratio remains
finite even in the chiral limit. However, as for the quantities
discussed above, finite volume corrections go in the direction of
invalidating this conclusion. Larger lattices are needed to draw
robust conclusions. 

%\begin{figure}[tp]
%\begin{center}
%\epsfig{file=gmor.eps,height=2.0in}
%\caption{The GMOR relation can be used to extract information on the
%  chiral condesate. The measure results however quite difficult in
%  practice and we cannot distinguish any signal of chiral symmetry
%  breaking.\label{fig:gmor}}
%\end{center}
%\end{figure}
The chiral condensate would be a prime candidate to study chiral
symmetry breaking. However due to the use of Wilson fermions, the
direct measure of $\langle\overline\psi\psi\rangle$ is plagued with UV
divergences which are notoriously difficult to tame.  Using the GMOR
relation an estimate for the chiral condensate can be
obtained\footnote{We do not attempt here to compute the necessary
  multiplicative renormalization constant, since we are not interested
  to the actual physical value. Perturbative results for the
  renormalization of fermions bilinears can be found in
  Ref.~\cite{DelDebbio:2008wb}.}. The method has been applied with
success in the case of QCD, see e.g. Ref.~\cite{Giusti:1998wy}.  We
present our result for this quantity in Fig.~\ref{fig:mvomps} (right).  Although
there is a partial cancellation of the finite size effects coming from
the pseudoscalar mass and the decay constant, the larger volume
dependence of the latter dominates yielding large systematic errors.
As a consequence an extrapolation is unfortunately not possible from
our current set of the data. We observe that finite volume effects
tend to make the condensate smaller, however the small numerical value
of the bare condesate by itself is not meaningful: for example in a typical QCD simulation
the value for this quantity is an order of magnitude smaller than the one
presented here.

\subsection{Discussion}
\label{sec:disc}
Lattice computations are a key ingredient for understanding strongly 
interacting gauge theories which may possess an IR fixed point.
To obtain reliable results
however, systematic errors must be kept under control. It is only in
recent years that the combination of increased computer power and more
efficient algorithms has allowed reliable predictions in the chiral
limit.  A lot of effort is required in order to reach the parameter
region of physical relevance for gauge theories with light quarks.
When considering theories other than QCD, lattice practitioners are
challenged by the lack of information available to guide them. In this
situations more than ever, sources of error should be kept under
control with great care.

% Consider the case of gauge theories which are candidate to possess an
% IR fixed point, such as $SU(3)$ with $N_f=12$ fundamental or $N_f=2$
% two-index symmetric Dirac fermions, or the case which we focused on
% $SU(2)$ with $N_f=2$ fermions in the adjoint representation.
% An IR fixed point can exist only with massless fermions. This is why
% many studies aiming to look for an IR fixed point use the
% Schr\"odinger functional
% formalism~\cite{Appelquist:2007hu,Shamir:2008pb,Hietanen:2009az},
% although only one of them~\cite{Appelquist:2007hu} attempt to
% perform a continuum extrapolation.

% The reason is clear when one consider than any such candidate theory
% has a perturbative $\beta$--function which is much smaller than in
% QCD. Any study aiming to measure this $\beta$--function, or better
% the continuum step scaling function, should be extraordinary
% accurate. What usually happen when one tries to perform such an
% extrapolation is that a small number compatible with zero within
% errors is obtained, but this does not help to resolve any IR fixed
% point in case it exists.

In this work we looked for evidence of an IR fixed point in the
spectrum of the theory. Several other
studies~\cite{Catterall:2007yx,Catterall:2008qk,DelDebbio:2008zf,Hietanen:2008mr,DeGrand:2008kx,Deuzeman:2009mh}
also investigate the light meson spectrum. However so far these
investigations have been limited to heavy mesons with masses of the
order of the UV cutoff, small volumes\footnote{There is only one work
  in the literature that uses lattices comparable to the ones used in
  this work~\cite{Hietanen:2008mr}, which however is limited to very
  heavy quark masses.} and, in some cases, at finite temperature.  All
these studies suffer from being in a parameter region in which the
systematic errors discussed in Sect.~\ref{sec:sim} mask the continuum
physics. Such investigations are useful to understand the phase
structure of the lattice theory and guide future investigations, for
example to avoid bulk transitions which are present in the lattice
formulation. However much more refined analyses are needed to extract
the interesting physics. Work in this direction is in progress in all
the groups involved in these studies. 

A first step in this direction was presented in this paper. We made
simulations at a fixed value of the lattice spacing, and aimed at
reaching the chiral limit in a controlled way.  Despite the use of
relatively large volumes up to $64\times 24^3$, finite size
corrections make the extrapolation to the chiral limit difficult. As
discussed in the previous section, the finite volume of the
system can mimic the existence of an IR fixed point in the chiral
limit. This makes the interpretation of our results quite difficult.
A possible way to understand if the behavior of the data presented in the last section
is already in the asymptotic region of small quark masses would 
be to identify a meaningful
reference quantity to fix the scale in the theory. In QCD it is
customary to use quantities such as the string tension $\sigma$ or the
Sommer scale $r_0$ in the chiral limit. For QCD, $r_0$ was chosen so
that $r_0 \approx 1/\sqrt\sigma$, and thus the two quantities stand on
equal footing.  If the theory if conformal in the IR however, the
string tension will vanish in the chiral limit, and it is therefore
not a good candidate for setting the scale. One can nonetheless use
$r_0$ as in QCD. In fact, one can always consider the static potential
between two fundamental charges in the theory $V(r)$ and the force
$F(r)\equiv dV(r)/dr$ associated with it. Then the
quantity $H(r)\equiv r^2 F(r)$ can be used to define the coupling
since it is proportional to $\alpha_{s}(\mu=1/r)$ in perturbation
theory.  In a QCD--like theory $H(r)$ will run from zero to infinity
and any value in this range can be used to set a scale
$r_0$. For QCD conventionally the Sommer scale is defined by $H(r_0)=1.65$.
If on the other hand, the massless theory is conformal in the IR, at
exactly zero quark mass $H(r)$ will run from zero to some finite value
$h^\star$. In this case any value in the range $]0,h^\star[$ can
be used to set a scale in the theory.  The presence of a finite
quark mass will spoil the IR conformal behavior. One would imagine
that at non-zero quark mass $m$ the coupling $H(x,m)$ will follow the
massless $H(x)$ up to some length scale $\tilde x(m)$; the latter
scale diverges as $m\rightarrow 0$, while at distances larger than
$\tilde x(m)$ the coupling will start to run again similar to what
happens in QCD. However if we choose a value $h\in]0,h^\star[$ to
define the scale $r_0$, such that $H(r_0,m)=h$, this quantity will
remain finite in the chiral limit, and could be used as a reference
scale.
\begin{figure}[tp]
\begin{center}
\epsfig{file=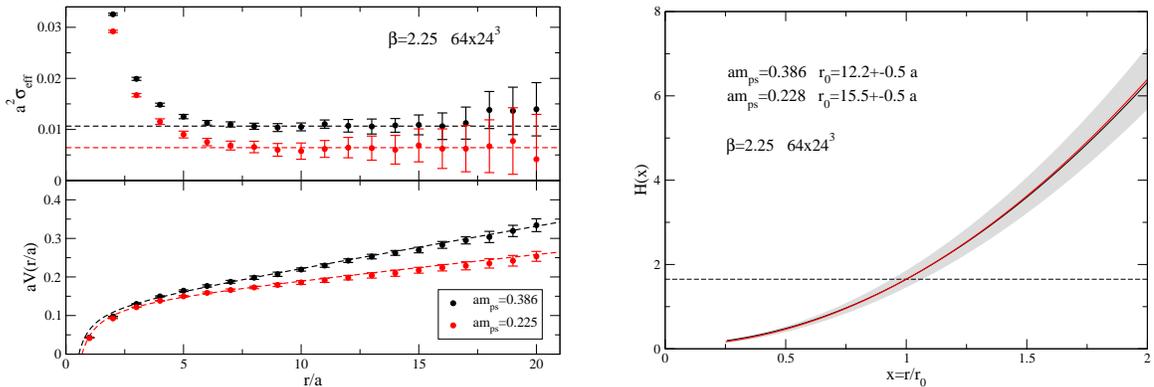,height=2.0in}\hfill
\epsfig{file=h.eps,height=2.0in}
\caption{(Left figure) upper panel: string tension as extracted from the Creutz
  ratios of Wilson loops for two different values of the quark
  mass. Lower panel: the corresponding static potential between
  fundamental charges extracted from Wilson loops. (Right figure) Evolution 
  of the coupling $H(r)=r^2F(r)$ in the $\overline q
  q$ scheme in units of $r_0$. The red curve corresponds to the
  lighter quark mass, while the virtually identical black line to the
  heavier mass.\label{fig:sp}}
\end{center}
\end{figure}
We show in Figs.~\ref{fig:sp} our preliminary results
for the static potential, the string tension and the coupling in the
$\overline q q$ scheme $H(r)$. These results are obtained on a
$64\times 24^3$ lattice at two of the smallest mass studied in this
work, and only a limited number of configurations, and as such should
be taken with a grain a salt. One step of HYP smearing was used to
improve the signal.  We extract the string tension from Creutz ratios
of Wilson loops, Fig.~\ref{fig:sp} (left) upper panel, which can then be used
as an input to fit the static potential $V(r)$ extracted from the
(rectangular) Wilson loops.  The parametrization used in the fit is
the usual Cornell potential: $V(r)=\alpha+\sigma r + \gamma/r$.
Using the fitted values for the string tension $\sigma$ and $\gamma$,
we can plot in Fig.~\ref{fig:sp} (right) the coupling $H(r)$, from which the
Sommer scale can be defined: $H(r_0)=1.65$. Here in the definition we
used the same value as in QCD, as the measured coupling does not show
a plateau in the limited range of distances explored in our data.

We measure a small string tension in units of the lattice spacing,
$a\sqrt\sigma\simeq 0.1$, and a quite large $r_0\simeq 15 a$. Similar
values for the string tension are also obtained when using a much more
refined variational method based on expectation values of Polyakov
loops (see Ref.~\cite{DelDebbio:2009fd} for a summary of our results;
a more detailed report is in preparation). Moreover, the dependence of
the string tension on the quark mass seems to be pronounced, while
$r_0$ appears to be more stable. The coupling $H(r)$, when plotted in
units of $r_0$ as in Fig.~\ref{fig:sp} (right), shows no residual dependence on
the mass within errors.
While the physical meaning of $r_0$ defined as above if the theory is conformal is not
clear, such large values of $r_0$ may suggest that the pseudoscalar
masses of order $\simeq 0.2 a^{-1}$ or bigger, are still ``heavy'' and
not in the chiral regime of the theory.  For example in QCD the values we observed, i.e.
$m_{\pi}r_0\approx 3.4$, corresponds to $m_{\pi}\approx 1.4$ GeV in a
box of $L\approx 0.8$ fm, and features similar to the ones we show in
this report are expected.
%  It is therefore questionable if one can
%conclude that the observed features of the spectrum are due to the IR
%conformal behavior of the theory.

% One way to determine if the theory is QCD-like or IR conformal, is
% to look at the theory in the regime where $m_\mathrm{PS}\ll
% 1/r_0$. This is a regime which is always possible to reach in both
% the scenarios we are considering. In that mass range the signature
% feature of the QCD-like theory, like a square root dependence of the
% pseudoscalar mass on the PCAC mass or a large splitting between the
% pseudoscalar and the vector mass, should be recovered, if this is
% indeed a QCD-like theory. In the IR conformal case instead, we
% should still observe a small numerical value for the ratio
% $m_{v}/m_\mathrm{PS}$ and the pseudoscalar mass large compared to
% $1/\sqrt\sigma$ and to the lowest glueball mass.

\section{Conclusions}
\label{sec:concl}
We have presented a careful investigation of the mesonic spectrum of
one of the prime candidate theories for a realistic technicolor
model. To study the low-lying spectrum of the SU(2) gauge theory with
two Dirac adjoint fermions we used numerical lattice simulations.
Such numerical simulations are an extremely powerful tool to explore
the non-perturbative dynamics of gauge theories which is otherwise
inaccessible to theoretical speculations, but great care must be taken
to control systematic errors.
% However as explained at length in Sect.~\ref{sec:sim}, numerical
% simulations are inevitably affected by systematic errors. The
% continuum physics which we aim to study, is only accessible after
% performing a number of extrapolations. For this reason it is
% essential to have control over these systematic errors in order to
% obtain reliable predictions. This is especially true for the theory
% under consideration for which our knowledge is seriously limited. To
% main objective of this study was to determine if the theory
% considered here lies within the conformal window or not.
We presented here the first study of the low-lying spectrum aiming at
the chiral limit in a controlled way. Using a series of four different
lattice sizes we showed how finite-size effects could be kept under
control in the range of quark masses explored.  The resulting behavior
of the different quantities analyzed, namely the pseudoscalar and
vector meson mass and the pseudoscalar decay constant, show
significant deviations from what is expected in the presence of
spontaneous chiral symmetry breaking.  In particular we showed that
the pseudoscalar mass does not seem to scale with the square root of
the quark mass, the ratio of the vector to the pseudoscalar mass
differs from one only by a few percents, and $F_\mathrm{PS}$ seems to
extrapolate to zero in the chiral limit.  While our data alone is
still not enough for a proof, if the behavior we observed persists
down to arbitrary small quark masses this will be a clear evidence for
the existence of an IR fixed point.

One of the major difficulties at present is to set a significant
reference scale for comparing other measured quantities. We showed
that a scale analogous to $r_0$ in QCD can always be defined even if
the theory is conformal, but in this case its physical meaning is not
clear. With the same definition as in QCD, the numerical value for
$r_0$ is around $15a$ at the smallest quark mass used in this study. We
also measured the string tension, obtaining quite small values in units
of the lattice spacing. This is certainly consistent with an IR fixed
point, but again not enough to draw any definitive conclusion.

To find solid evidence for the existence of an IR fixed point will
certainly require a large numerical effort. The use of different and
complementary approaches to the one presented here, like the study of
gluonic observables, or the measure of the running couplings to study the
renormalization group flow, will be beneficial.

\end{document}